\documentclass{moriond}

\usepackage{amsmath,amssymb,amsfonts}
\usepackage{floatrow}

\def\be{\begin{equation}}
\def\ee{\end{equation}}
\def\bea{\begin{eqnarray}}
\def\eea{\end{eqnarray}}

\newcommand{\Photo}{}

\begin{document}
\begin{flushright}
\texttt{DESY-26-053}
\end{flushright}
\vspace*{2cm}
\title{Precision predictions for trilinear scalar couplings\\ and Higgs pair production in models with extended scalar sectors}

\author{J. Braathen}

\address{Deutsches Elektronen-Synchrotron DESY, Notkestr.~85,  22607  Hamburg,  Germany}

\maketitle\abstracts{
Reconstructing the shape of the Higgs potential realised in Nature is one of the most pressing tasks of current and future colliders. This will offer deep insights into the dynamics of the electroweak phase transition and provide a unique opportunity to probe physics beyond the Standard Model (BSM). In this context, it is essential to have precise theory predictions for trilinear scalar couplings, which control the form of the potential, and for Higgs pair production processes, which are the observables that allow accessing the trilinear couplings. 
I summarise in these proceedings recent progress in precision calculations of both trilinear scalar couplings and Higgs pair production at the (HL-)LHC in BSM models with extended scalar sectors. 
}

\section{Introduction}
Probing the shape of the Higgs potential that is realised in Nature is essential to understand the dynamics of the electroweak phase transition, and search for signs of Beyond-the-Standard-Model (BSM) physics. A crucial quantity in this context is the trilinear self-coupling of the detected Higgs boson, $\lambda_{hhh}$, which controls the form of the potential along the Higgs field direction. 
In many theories with extended scalar sectors, radiative corrections from the BSM scalars can give rise to sizeable deviations in $\lambda_{hhh}$ from its Standard Model (SM) value, see e.g. Refs.~\cite{Kanemura:2004mg,Braathen:2019pxr,Braathen:2019zoh,Bahl:2022jnx,Bahl:2023eau,Basler:2024aaf} and references therein. 
Moreover, the comparison of high-precision predictions for $\lambda_{hhh}$ with the experimental limits derived from di-Higgs production searches --- the best current limits~\cite{CMS:2026nuu} being $-0.71<\kappa_\lambda\equiv \lambda_{hhh}/\lambda_{hhh}^{\text{SM},\ (0)}<6.1$ at 95\% C.L., where $\lambda_{hhh}^{\text{SM},\ (0)}$ is the tree-level SM prediction --- already provides a powerful tool to probe the parameter space of BSM theories, beyond the reach of other state-of-the-art experimental and theoretical constraints~\cite{Bahl:2022jnx}.
In this context, the inclusion of the known and numerically significant two-loop corrections to $\lambda_{hhh}$ is necessary for reliable interpretations of the experimental bounds in terms of the allowed or excluded regions of BSM parameter space --- thus motivating obtaining precision predictions for $\lambda_{hhh}$.  
For the purpose of reconstructing the potential of an extended Higgs sector, more trilinear scalar couplings --- generically denoted $\lambda_{ijk}$ --- are needed. These can appear in pair production of the detected Higgs boson, like $gg\to hh$ at the (HL-)LHC, or in BSM Higgs pair production processes. In general, $\lambda_{ijk}$ can also receive significant radiative corrections~\cite{Arco:2025pgx,Braathen:2025qxf}. 

In these proceedings, I summarise recent progress in the calculations of $\lambda_{ijk}$ and of Higgs pair production in models with extended scalar sectors, and the automation of these calculations.  

\section{Trilinear scalar couplings}
A first step towards automated calculations to probe the Higgs potential was taken with the first version of the public tool \texttt{anyH3}~\cite{Bahl:2023eau}, part of the broader framework \texttt{anyBSM}. \texttt{anyH3} (v1) enables the full one-loop calculation of $\lambda_{hhh}$ in any model defined by \texttt{UFO} model input files. Recently, the code has been extended~\cite{Bahl:2026nsu} to allow computing any trilinear coupling $\lambda_{ijk}$ at one loop.  
Additionally, complete two-loop results for the calculation of trilinear and quartic self-couplings of arbitrary scalars (i.e.\ $\lambda_{\phi_i\phi_i\phi_i}$ and $\lambda_{\phi_i\phi_i\phi_i\phi_i}$ where $\phi_i$ can be the detected Higgs boson $h$, or any BSM scalar) have also been derived for general renormalisable theories~\cite{Bahl:2025wzj}. Specifically, expressions have been obtained for genuine two-loop and subloop renormalisation contributions. Extensive use was made of symmetry properties of the considered diagrams, to simplify both the calculation and the application of the results to concrete models. 
Lastly, effective field theory techniques can also be applied in scenarios with large BSM deviations in order to incorporate numerically significant classes of higher-order corrections, even beyond two loops~\cite{Naskar:2026abc}.

\section{Automating predictions for Higgs pair production with \texttt{anyHH}}

The actual observables that offer access to trilinear scalar couplings and thus allow probing the shape of the Higgs potential are Higgs pair production processes, such as $gg\to h_i h_j$. 
Studies of the pair production of the detected Higgs boson, namely $gg\to hh$, at the (HL-)LHC in models with extended scalar sectors have received considerable attention in recent years.

\begin{figure}
    \centering
    \includegraphics[width=\textwidth]{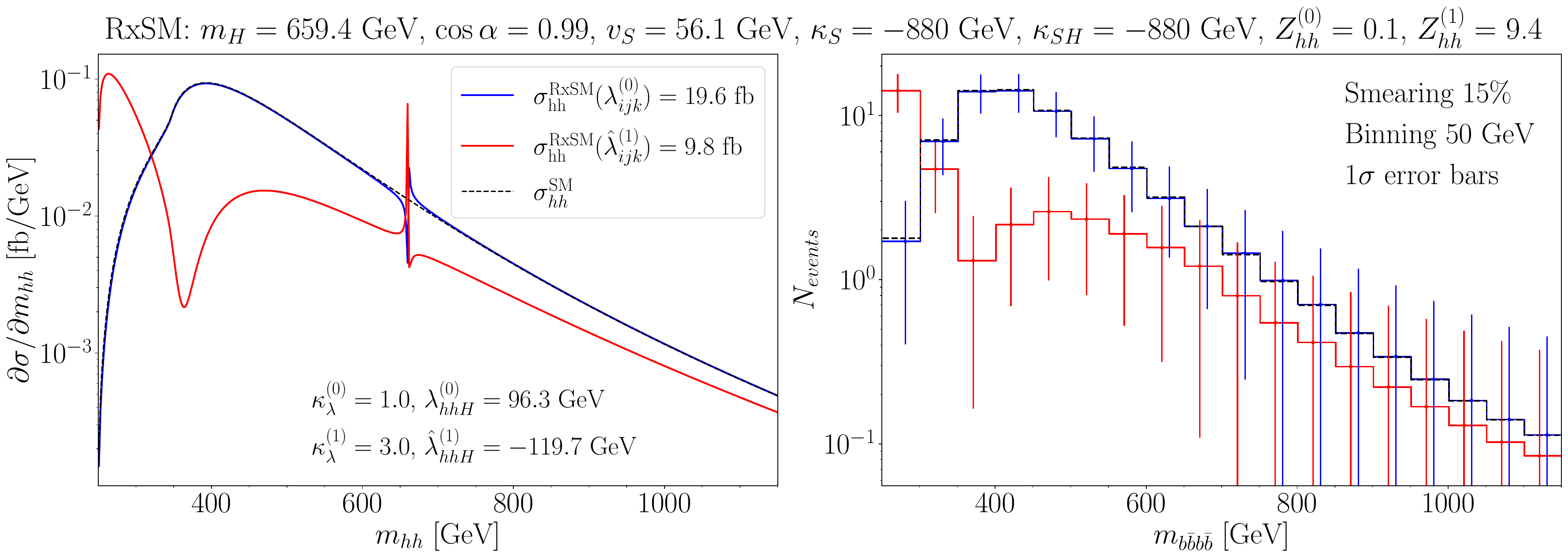}
    \caption{
\textit{Left:} Differential cross-section distributions in the RxSM without experimental uncertainties;
\textit{right:} distributions of the number of di-Higgs events in the $b\bar bb\bar b$ final state, taking into account 
smearing, binning and experimental efficiencies. Error bars correspond to Poisson statistical uncertainties for each bin. Blue (red) curves indicate results using tree-level (one-loop) values for $\lambda_{hhh}$ and $\lambda_{hhH}$, while the black dashed line is the SM result.
    }
    \label{fig:RxSM_hh}
\end{figure}

Several of these works~\cite{Heinemeyer:2024hxa,Feuerstake:2024uxs,Arco:2025nii} have highlighted the importance of the interference between resonant and non-resonant contributions to this process, and the need to take this into account when interpreting experimental search limits. 
Moreover, it was also found that the inclusion of loop corrections to the trilinear scalar couplings appearing in the diagrams contributing to Higgs pair production can have a drastic impact on total and differential cross-sections~\cite{Heinemeyer:2024hxa,Braathen:2025qxf} --- especially in scenarios in or near the alignment limit. An example~\cite{Braathen:2025qxf} of this is provided in Fig.~\ref{fig:RxSM_hh} for a scenario of the general singlet extension of the SM, the RxSM. The left plot shows theoretical curves for the differential cross-section distributions as a function of the di-Higgs invariant mass $m_{hh}$, while in the right plot event numbers are shown after experimental effects and statistical errors have been included (see details in the caption and in Refs.~\cite{Arco:2025nii,Braathen:2025qxf}). The blue (red) lines are obtained using tree-level (one-loop) values for the trilinear scalar couplings in the di-Higgs calculation, and the black line indicates the SM result.\footnote{In the SM, the loop corrections to $\kappa_\lambda$ are moderate, hence there is no significant difference between SM curves obtained with tree-level or one-loop trilinear couplings.}
The benchmark scenario considered here (defined in the title of Fig.~\ref{fig:RxSM_hh}) shows that the inclusion of one-loop corrections to $\lambda_{ijk}$ can have a drastic impact on the $m_{hh}$ distributions. 
The large BSM effects in $\kappa_\lambda$ significantly modify the interference pattern between the box and SM-like triangle diagrams near the di-Higgs production threshold at $m_{hh}\gtrsim 250\text{ GeV}$. Furthermore, the loop corrections flip the sign of the $\lambda_{hhH}$ coupling, which in turn leads to a change in the structure of the resonance around $m_{hh}=m_H$, from peak-dip to dip-peak. Additionally, a statistical significance $Z_{hh}$ to distinguish the RxSM from the SM was defined, taking into account experimental effects and Poisson statistical errors. Conclusions about whether the RxSM scenario in Fig.~\ref{fig:RxSM_hh} could be distinguished from the SM would change entirely depending on the order of the $\lambda_{ijk}$ couplings used the analysis: $Z_{hh}^{(0)}=0.1$ with tree-level values --- i.e.\ there appears to be no sensitivity to detect BSM physics --- while $Z_{hh}^{(1)}=9.4$, well above the $5\sigma$ threshold to claim a discovery, when using loop-corrected couplings.

In order to provide precise predictions for Higgs pair production in arbitrary models with extended scalar sectors, the \texttt{anyBSM} framework has been extended with \texttt{anyHH}~\cite{Bahl:2026nsu}. This new module enables the calculation of total cross-sections and differential $m_{hh}$ distributions for processes $gg\to h_ih_j$, where $h_ih_j$ can be any pair of (potentially non-identical) neutral scalars.
In particular, \texttt{anyHH} allows including \textit{(i)} all non-resonant and resonant contributions (for any number of resonant scalars); \textit{(ii)} complete one-loop corrections to all relevant trilinear scalar couplings; \textit{(iii)} one-loop propagator corrections in $s$-channel diagrams; and \textit{(iv)} the full momentum-dependence of the trilinear couplings. \texttt{anyHH} requires inputs in the \texttt{UFO} format, like \texttt{anyH3}, and is so far restricted to models without additional coloured particles.  

\begin{figure}
    \centering
    \includegraphics[width=\textwidth]{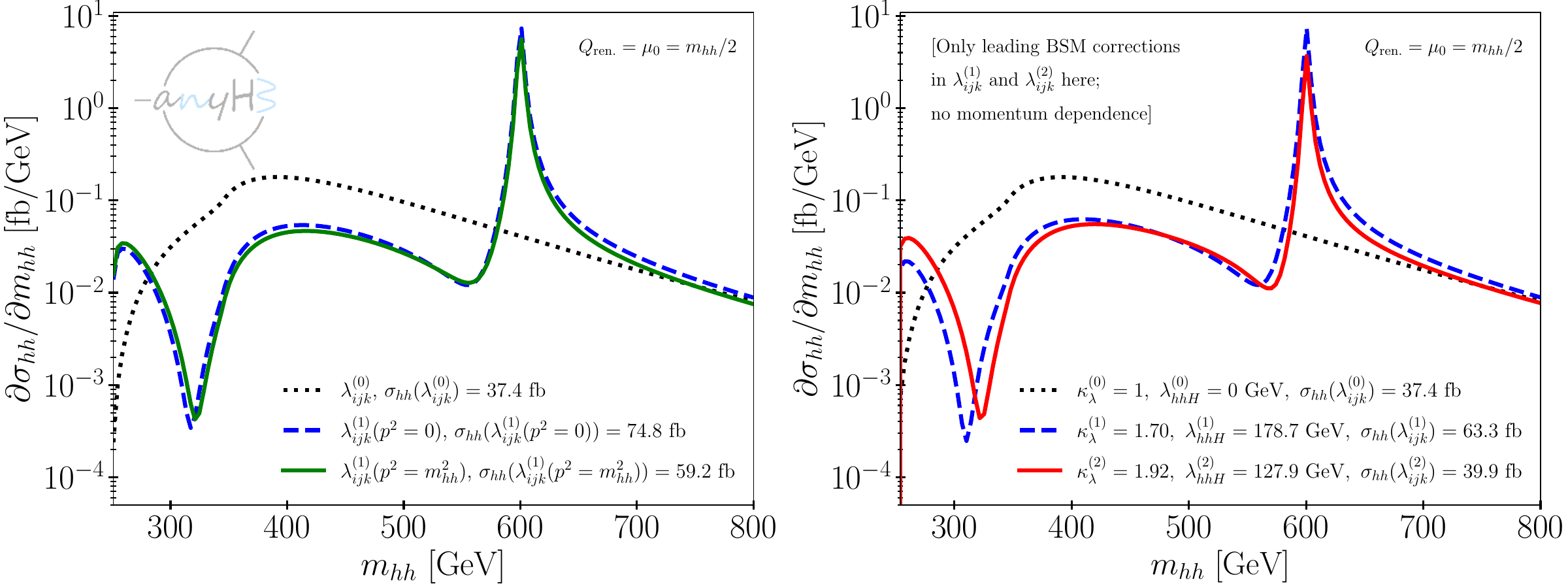}
    \caption{Differential $m_{hh}$ distributions obtained with \texttt{anyHH}, for a 2HDM scenario in the alignment limit ($\alpha=\beta-\pi/2$) and with $M=400\ \mathrm{GeV}$, $m_H=m_A=m_{H^\pm}=600\ \mathrm{GeV}$ and $t_\beta=2$. A QCD $K$-factor of 2 is used to approximate the contribution of NLO QCD corrections.  
    \textit{Left:} with the in-built options of \texttt{anyHH}, i.e.\ using tree-level (black dotted line) or one-loop trilinear couplings with (blue dashed) or without (green solid) momentum dependence; \textit{right:} including also leading two-loop corrections to the trilinear couplings (red solid line). 
    The one-loop results differ slightly between the two plots as only the leading one-loop effects are included in the right plot and because of differences in the choices of renormalisation schemes for the calculation of $\lambda_{ijk}$.}
    \label{fig:anyHH}
\end{figure}

Differential $m_{hh}$ distributions obtained with \texttt{anyHH} for a scenario of the Two-Higgs-Doublet Model (2HDM) are displayed in the left plot of Fig.~\ref{fig:anyHH} 
(for details on the model and notations see Ref.~\cite{Bahl:2023eau}). 
The scenario chosen for the left plot is in the alignment limit, meaning that at the tree level $\kappa_\lambda^{(0)}=1$ and $\lambda_{hhH}^{(0)}=0$. Turning on loop corrections then yields substantial changes to both couplings, with one-loop values of $\kappa_\lambda^{(1)}\simeq 1.8$ and $\lambda_{hhH}^{(1)}\simeq 192\ \mathrm{GeV}$, which in turn drastically modifies the $m_{hh}$ distributions. The increase in $\kappa_\lambda$ alters the interference pattern between box and triangle contributions and drives the location of maximal interference towards higher $m_{hh}$. Meanwhile the loop-induced value of $\lambda_{hhH}$ implies that a resonant peak appears around $m_{hh}\simeq m_H$ once one-loop trilinear couplings are considered. Both effects have a strong impact on the total cross-section as well, which is approximately doubled. Comparing the blue and green curves, one finds that the inclusion of the momentum dependence does not affect the shape of the distributions very much but leads to a reduction of $\sim 20\%$ of the total cross-section.

Finally, going beyond the one-loop level, the impact of the dominant two-loop corrections to the trilinear couplings on Higgs pair production has also been considered recently~\cite{Degrassi:2025pqt,Braathen:2026glt}. While the inclusion of leading two-loop BSM effects does not modify the shape of differential $m_{hh}$ distributions~\cite{Braathen:2026glt} qualitatively, they can still have a significant numerical impact, as illustrated in the right plot of Fig.~\ref{fig:anyHH} (where two-loop results for $\lambda_{hhh}$ and $\lambda_{hhH}$ have been interfaced with \texttt{anyHH}). 
Most significantly, the two-loop corrections further increase the BSM deviation in $\kappa_\lambda$, which shifts the location of the destructive interference between box and triangle contributions further away from the threshold at $\sim250\ \mathrm{GeV}$. The width and height of the resonant peak are moreover affected by the $\sim 30\%$ decrease in $\lambda_{hhH}$ from the one- to two-loop order. The combined effects on the interference pattern near the threshold and the change of the resonant peak leads to a reduction of the total cross-section of $\sim 37\%$.

Examples of new results obtained with \texttt{anyHH}, in models for which no other public tool for di-Higgs production is available or in scenarios with multiple resonances can be found in Ref.~\cite{Bahl:2026nsu}.

\section{Conclusions}
I have briefly presented here recent results for calculations of trilinear scalar couplings $\lambda_{ijk}$ and for Higgs pair production at the (HL-)LHC, with a particular emphasis on the public tools \texttt{anyH3} and \texttt{anyHH}. I also provided benchmark scenarios illustrating the strong impact that BSM radiative corrections to $\lambda_{ijk}$ can have on di-Higgs predictions. While these examples clearly exhibit the desired perturbative convergence of the calculations, they demonstrate the importance of including higher-order BSM corrections in precision predictions for Higgs pair production.

\section*{Acknowledgments\vspace{-2pt}}
I would like to thank the Moriond organisers for the kind invitation. I am also deeply grateful to all my collaborators for our work together over the years. 
I am supported by the DFG Emmy Noether Grant No.\ BR 6995/1-1.
I also acknowledge support by the Deutsche Forschungsgemeinschaft (DFG, German Research Foundation) under Germany's Excellence Strategy --- EXC 2121 ``Quantum Universe'' --- 390833306. This work has been partially funded by the Deutsche Forschungsgemeinschaft (DFG, German Research Foundation) --- 491245950.

\end{document}